\documentclass{emulateapj}

\shorttitle{The period analysis of V418 Aql, SU Boo, RV CVn, CR Cas, GV Cyg, V432 Per, and BD+42
2782.} \shortauthors{Zasche et al.}

\begin{document}

\title{The period analysis \\ of V418 Aql, SU Boo, RV CVn, CR Cas, GV Cyg, V432 Per, and BD+42
2782.}

\author{P. Zasche\altaffilmark{1} and M. Wolf\altaffilmark{1} and R. Uhla\v{r}\altaffilmark{2} and H. Ku\v{c}\'akov\'a\altaffilmark{1,3}}
\affil{
 \altaffilmark{1} Astronomical Institute, Charles University in Prague, Faculty of
Mathematics and Physics, CZ-180 00 Praha 8, V Hole\v{s}ovi\v{c}k\'ach 2, Czech Republic \\
 \altaffilmark{2} Private Observatory, Poho\v{r}\'{\i} 71, CZ-254 01 J\'{\i}lov\'e u Prahy, Czech
Republic \\
 \altaffilmark{3} Johann Palisa Observatory and Planetarium, Technical University Ostrava,
CZ-708 33 Ostrava, Czech Republic \\}

\begin{abstract}
\noindent The minimum timings of eclipsing binaries V418 Aql, SU Boo, RV CVn, CR Cas, GV Cyg, V432
Per, and BD+42 2782 were collected and analyzed. Their long-term behavior was studied via period
analysis, revealing a periodic term in eclipse times. We derived 576 new times of minimum. Hence,
to describe the periodic variation, a thirdbody hypothesis was proposed and the resulting orbital
periods are as follows: 70, 7.4, 53, 37, 27, 53, and 18 yr, respectively. For the system V432 Per
an additional 9.5 yr variation was also found. The predicted minimum masses of these distant bodies
were calculated and their detectability discussed. The light curves of SU Boo and RV CVn were
analyzed using the PHOEBE program, resulting in physical parameters of the components. New variable
stars in the field of V418 Aql were discovered.
\end{abstract}


\section{Introduction}

More than a century of intensive study of the eclipsing binaries (hereafter EBs), these objects
still represent the best method how to derive the masses, radii and luminosities of the stars.
Moreover, discovering the additional components in these systems is also rather straightforward
using the precise times of minima and analysing the period variation of the eclipsing pair, a
so-called light-time effect (hereafter LITE), \cite{Irwin1959} or \cite{Mayer1990}.

The method of period analysis, despite its classical nature and many decades of usage (about 250
such systems known nowadays, see e.g. \citealt{2010KPCB...26..269Z}), still provides us with an
efficient method of discovering the hidden components in the eclipsing systems. Its main advantage
is the easiness of use because of huge datasets of times of eclipses. The other profit is that this
method is able to reveal the hidden components otherwise hardly detectable: the short-periodic ones
can be easily detected via spectroscopy, while the long-period ones as visual or interferometric
doubles. Hence, the period gap in between can be harvested via period analysis in these systems -
it is adequately sensitive to relatively low masses, independent of luminosities of the third
bodies and also only mildly dependent on the orbit orientations (only the body orbiting
perpendicular to the EB orbit cannot be detected). And finally, its usefulness also for the huge
photometric databases was shown (e.g. by \citealt{2013ApJ...768...33R}).

\section{Methods}

Only for a brief repetition of the method of period analysis using the LITE hypothesis:
\begin{equation} \tau = \frac{A} {\sqrt{1-e^2\cos^2\omega}} \left[ {{
\frac{(1\!-\!e^2) \!\cdot \sin(\nu + \omega)}{1+e \cos\nu} + e \sin\omega}} \right] \end{equation}
is the delay of the light-time orbit as the body moves around a common barycenter (see e.g.
\cite{Mayer1990} for explanation of the individual parameters). This delay is periodically changing
with respect to the current orbital phase, hence the times of minima for a particular system are
being observed earlier and later than predicted from the linear ephemerides. For some of the
systems also the quadratic term in ephemerides was used. Hence another parameter of a rate of
period change $q$ was introduced. This continuous period change is often attributed to the mass
transfer between the close eclipsing components. The mass transfer is slowly moving the barycenter
of the double and hence also the period of the pair itself. Using the hypothesis of a conservative
mass transfer (i.e. no mass loss from the system), the well-known equation introduced e.g. by
\cite{2001icbs.book.....H} can be used to compute the estimated rate of the mass transfer:
\begin{equation} \frac{1}{P} \frac{\mathrm{d} P}{\mathrm{d} t} = 3 \, \frac{M_1\!-\!M_2}{M_1 M_2}
\, \frac{\mathrm{d} M_1}{\mathrm{d} t}, \label{masstransfer}\end{equation} where $M_i$ are the
masses of primary and secondary component, respectively.

There are still many eclipsing systems lacking of detailed period analysis despite the fact that
their observations exist in various databases. For example the automatic photometric projects (like
ASAS \citep{2002AcA....52..397P}, Super WASP \citep{2006PASP..118.1407P}, "Pi of the sky"
\citep{2005NewA...10..409B}, NSVS \citep{NSVS}, OMC \citep{OMC} and others) monitor the sky
continuously, and the data are publicly available. These data points can be used either for
deriving the times of minima, or for the complete light curve (hereafter LC) analysis. For our
study we have chosen several rather neglected eclipsing binaries on the northern sky for their
availability from our observatories.

\section{New photometric observations}

New observations were mostly obtained at Ond\v{r}ejov observatory in Czech Republic, using the
65-cm reflector equipped with the MI G2-3200 CCD camera. The standard $R$ photometric filter was
used, while the exposing times were chosen according to the brightness of the target (usually 10-90
s). The only exception was the star BD+42 2782, which is too bright for this telescope, hence is
was observed by one of the authors (RU) with small 34-mm and 200-mm telescopes at the private
observatory in J\'{\i}lov\'e u Prahy in Czech Republic. The observations were obtained using the
standard $R$ filter, or without any filter.

All of the observations were routinely reduced in a standard way, using the dark frames and flat
fields. The resulting photometry was used for deriving the times of minima for a particular system.
The standard Kwee-van Woerden procedure \citep{Kwee} was used for the derivation of the times of
minima. And finally, the heliocentric correction was applied to the data points. All of the data
points used for the analysis are stored in the Appendix Tables below. All of these times of minima
are the heliocentric ones (HJD). Also the accuracy of the particular minima are given in the
tables, for our new derived ones as well as for the already published ones (if available).

\section{The individual systems under analysis}

In the present analysis we included only such systems, which satisfy all of the following criteria:
 \begin{itemize}
   \item Northern-sky eclipsing binary in the range 9 -- 15 mag and orbital period up to 3 days\\[-6mm]
   \item The times-of-minima data set is sufficiently large for a period analysis\\[-6mm]
   \item The variation in the $O-C$ diagram shows periodic variation and at least one period of
   such variation is covered nowadays\\[-6mm]
   \item The system was not studied before, or a new solution significantly differs from the
   published one\\[-6mm]
   \item At least a few new minima times observations were obtained by the authors during the last
   years\\[-6mm]
 \end{itemize}
Using these criteria, seven systems were found to be suitable for the analysis.

\subsection{V418 Aql}

V418 Aql (= AN 115.1930, $V$=11.6~mag) was discovered as a variable star by
\cite{1939AN....268..165G}, who also correctly classified the star as an Algol-type. However, since
then only a few studies on this star was published, and no detailed photometric or spectroscopic
study was performed. The spectral type was classified as F8III \citep{1984PASP...96...98H}, but
based on only fair quality of spectrograms. Later, \cite{1987BBSAG..84Q...6L} published his finding
on the duration of the totality of the primary eclipse of about 2 hours, which is only a bit longer
than derived from our new observations ($1^\mathrm{h}43^\mathrm{m}$). Moreover, the system V418~Aql
comprises two components, see the Washington double star catalogue, WDS \citep{WDS}. The secondary
component is about 17$^{\prime\prime}$ distant.

\begin{figure}
  \centering
  \includegraphics[width=0.48\textwidth]{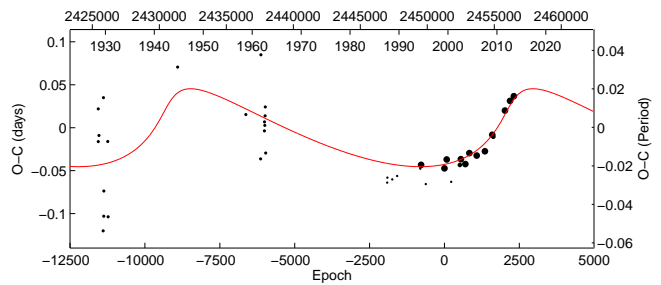}
  \caption{Period analysis of V418 Aql. The individual times of minima are plotted as dots, bigger the symbol,
  higher the weight, while the continuous curve represents the final fit, see the text for details.}
  \label{FigV418Aql_LITE}
\end{figure}

We collected all available times of minima of V418~Aql, see the Appendix Table \ref{TabMIN}. For
deriving new times of minima we also used the public available photometry obtained for the ASAS
survey \citep{2002AcA....52..397P}, NSVS survey, and the OMC camera onboard the INTEGRAL satellite.
Two new minima were also observed during the last year by the authors. The data were analysed
applying the LITE hypothesis. See Fig. \ref{FigV418Aql_LITE} for the final result, the parameters
of the LITE orbit are given in Table \ref{LITEparam}. In this table there are presented all the
fitted parameters from the LITE hypothesis together with their respective errors. However, it is
necessary to emphasize that these errors are only formal errors as resulted from the fitting
procedure (for the estimation of errors from the covariance matrix see e.g.
\citealt{NumRecipes1986}). Hence, these mathematical errors can sometimes be 2 to 5 times lower
than more reliable physical uncertainties of the individual parameters. As one can see, the
periodic variation is clearly visible, its period is about 70 years, while the periastron passage
will occur in upcoming years. Hence, new observations would be very welcome.

One can argue, that the whole analysis and our solution is based on one crucial point only, the one
near the last periastron in 1944. However, this is not true. We tried to perform a similar analysis
using only the data after 1950, and using only the quadratic term in ephemerides with no LITE
variation. But this approach did not led to the better result due to the fact that the curvature of
the points is not symmetric for a parabola and some of the points significantly deviate.

 \begin{figure*}
  \centering
  \includegraphics[width=0.96\textwidth]{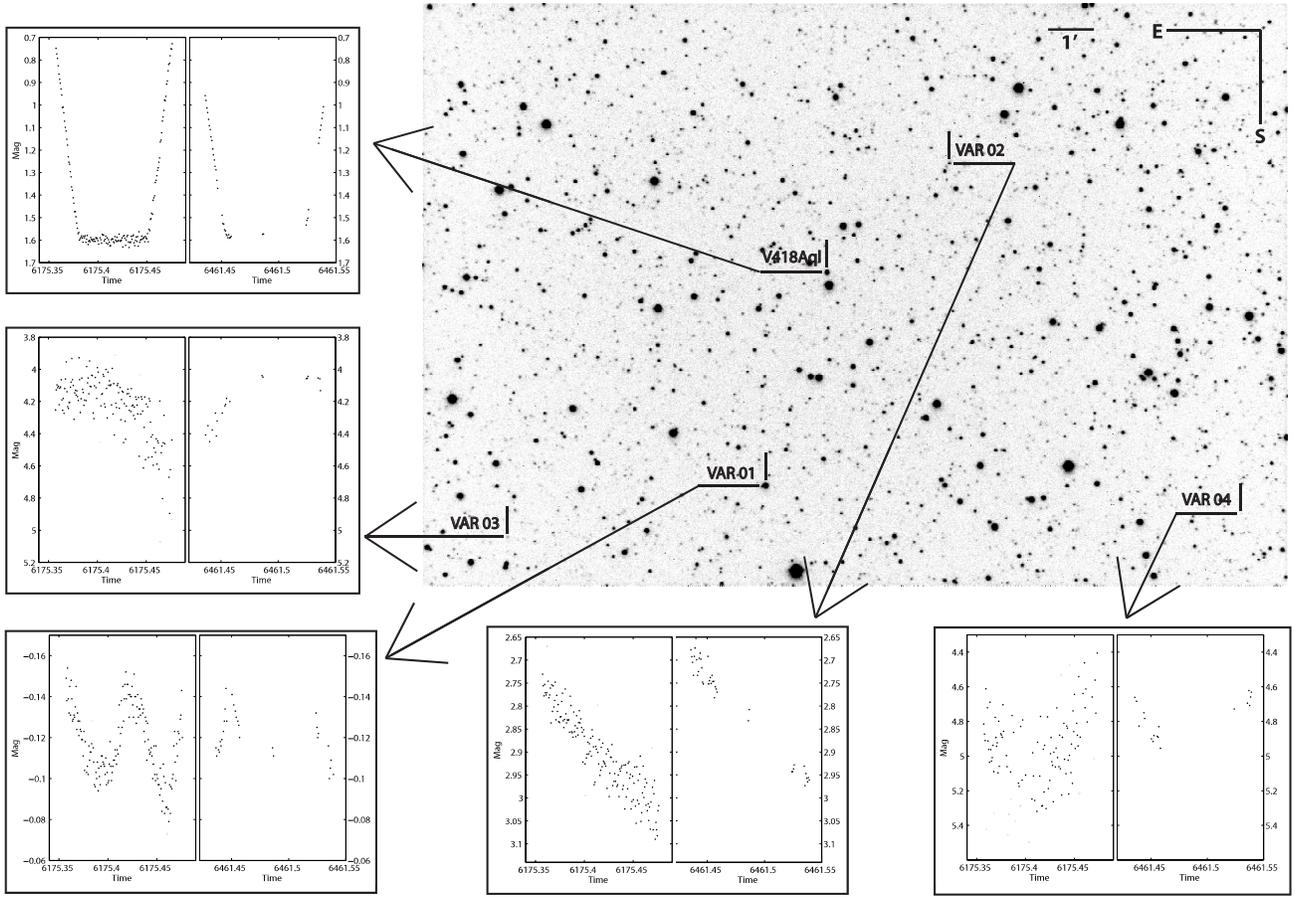}  
  \caption{Identification chart for the field of V418 Aql and the surrounding new variables, see the text for details.}
  \label{FigV418Aql_identNewVAR}
\end{figure*}

From the LITE hypothesis there arises that the mass function of the third hidden body is about 0.2
M$_\odot$, hence one can speculate about its nature. Using the easiest assumption about the
coplanarity of the both orbits (and using the total mass of the eclipsing pair about 1.6
M$_\odot$), there resulted the minimal mass of the third body about 1.1 M$_\odot$. Such a component
should be easily detectable in the light curve solution and should be also visible in the spectra
of the system. New detailed analysis is hence needed. And finally, there also arises that this
third component is different than the one observed visually, hence we deal with at least a
quadruple stellar system.

One can also ask whether such a picture of the system is self-consistent with the individual
luminosities. Using the spectral type F8III as derived by \cite{1984PASP...96...98H}, its absolute
bolometric magnitude is about 3 mag brighter than the normal main sequence F8 stars (see e.g.
\citealt{2000asqu.book.....C}). About the same spectral type was also derived using the photometric
indices ($V-K$) and ($J-H$) of V418~Aql observed by 2MASS survey \citep{2006AJ....131.1163S}.
According to our observations, the primary minimum is about 2.36~mag deep, while the secondary
about 0.04~mag only. Hence, the primary giant component contributes about 90\% of the total
luminosity of the system and is the absolutely dominant source. This is also the reason, why the
observed combined spectral type F8III is mainly the spectral type of the primary component. We
tried to find out the individual properties of the three components in the system from our (poorly
covered) light curve. There resulted that the secondary is probably a subgiant of spectral type
about M1IV. Hence, the total mass of the eclipsing binary is about 1.2 + 0.4 = 1.6~$M_\odot$. From
the fitting procedure there was also derived that the value of the third light is about 4\% of the
total light. The light contribution of the third body with mass 1.1~$M_\odot$ (i.e. about G5V
spectral type) as derived from the LITE analysis is about 3.5\%, which is in excellent agreement.

Moreover, during the observations of V418 Aql we discovered several new variable stars in the
field. See Fig. \ref{FigV418Aql_identNewVAR} for the identification chart and position of the new
variables near V418~Aql. Our two nights of observations are plotted for each of these stars. The
brightest one (designated as VAR 01) is the star GSC 0048604545 (= 2MASS 19364467 +0352167, RA
19$^h\!$ 36$^m\!$ 44$^s\!\!$.70, DE $+$03$^\circ\!$ 52$^{\prime}\!$ 16$^{\prime\prime}\!\!$.3). It
is a rapidly pulsating star, probably of $\delta$ Scuti type. Its variations are of about 0.07~mag
in $R$ filter, while the period of pulsations is about 1.5~hours. The second star (VAR 02) is 2MASS
19362870 +0359267 (RA 19$^h\!$ 36$^m\!$ 28$^s\!\!$.73, DE $+$03$^\circ\!$ 59$^{\prime}\!$
27$^{\prime\prime}\!\!$.0), but its type is unknown, having the amplitude at least 0.35~mag. The
two other new variables (VAR 03 = 2MASS 19370740 +0351051, RA 19$^h\!$ 37$^m\!$ 07$^s\!\!$.40, DE
$+$03$^\circ\!$ 51$^{\prime}\!$ 05$^{\prime\prime}\!\!$.17 and VAR 04 = 2MASS 19360258 +0351466, RA
19$^h\!$ 36$^m\!$ 02$^s\!\!$.58, DE $+$03$^\circ\!$ 51$^{\prime}\!$ 46$^{\prime\prime}\!\!$.64) are
rather faint, but their variations are still visible in the data, see Fig.
\ref{FigV418Aql_identNewVAR}.

\begin{table*}
 \centering
 \tiny
 \begin{minipage}{230mm}
  \caption{Final parameters of the LITE orbits.}  \label{LITEparam}
  \begin{tabular}{@{}c c c c c c c@{}}
\hline
 Parameter           &   V418 Aql        &  SU Boo           &  RV CVn          &  CR Cas           &  GV Cyg           &  BD +42 2782   \\
 \hline
 $JD_0 $             & 2451276.4853 (106)& 2453142.5733 (18) & 2444374.6415 (16)& 2440529.0619 (96) & 2450283.4500 (45) & 2444423.3372 (19) \\
 $P$ [d]             & 2.23490129 (168)  & 1.56125039 (20)   & 0.26956736 (3)   & 2.84019694 (262)  & 0.99066628 (56)   & 0.37015161 (9)    \\
 $p_3$ [day]         & 25548.4 (953.1)   & 2709.1 (24.8)     & 19397.8 (852.3)  & 13553.0 (789.6)   & 9847.1 (442.9)    & 6470.60 (67.49)   \\
 $p_3$ [yr]          & 69.95 (2.61)      & 7.42 (0.07)       & 53.1 (2.3)       & 37.1 (2.2)        & 27.0 (1.2)        & 17.7 (0.2)        \\
 $A$  [day]          & 0.0453 (96)       & 0.0076 (5)        & 0.0074 (9)       & 0.0451 (32)       & 0.0079 (8)        & 0.0099 (5)        \\
 $T_0$               & 2430626.6 (760.0) & --                & 2453523.1 (892.0)& --                & --                & 2490833.6 (133.6) \\
 $\omega$ [deg]      & 27.9 (15.8)       & --                & 131.5 (31.9)     & --                & --                & 85.5 (6.8)        \\
 $e$                 & 0.658 (0.305)     & 0.000 (0.001)     & 0.413 (0.014)    & 0.000 (0.001)     & 0.000 (0.001)     & 0.546 (0.090)     \\
 $q$ [10$^{-10}$ d]  & --                & 1.695 (0.001)     & 0.024 (0.001)    & 24.04 (0.01)      & 0.693 (0.003)     & --                \\ \hline
 $f(m_3)$ [M$_\odot$]& 0.184 (56)        & 0.045 (0.002)     & 0.001 (0.001)    & 0.347 (0.015)     & 0.004 (0.001)     & 0.016 (0.001)     \\
 $M_{3,min}$ [M$_\odot$]& 1.1 (0.3)      & 0.97 (0.05)       & 0.17 (0.07)      & 6.63 (0.89)       & 0.29 (0.08)       & 0.51 (0.05)       \\
 $\dot{M}$ [M$_\odot$/yr]& --           &1.8 $\cdot 10^{-7}$&9.9 $\cdot 10^{-8}$&3.2 $\cdot 10^{-6}$&9.4 $\cdot 10^{-9}$& --                \\
 \hline
\end{tabular}
\end{minipage}
\end{table*}

\subsection{SU Boo}

Another star in our sample is SU Boo (= AN 78.1914, $V$=11.9~mag). It was discovered as a variable
by \cite{1914AN....198..371B}, while later \cite{1960MmSAI..31..107B} performed the first analysis
of its light curve. It is the classical Algol-type binary with deep primary and shallow secondary
minima, orbital period is of about 1.5~days, and the spectral type was derived to be A3V/A4V
\citep{1975MmRAS..79..131H}. The inclination of the system is about 81.5$^\circ$ according to
\cite{1960MmSAI..31..107B}, however later \cite{1980A.AS...40...57M} published the value $i =
86.3^\circ$. Therefore, such a large discrepancy is noteworthy and new LC analysis would be
profitable.

\begin{table}
 \centering
 \begin{minipage}{75mm}
  \centering
  \caption{The parameters of the light curves of SU Boo and RV CVn as derived from the analysis.}  \label{TabLC}
  \begin{tabular}{@{}c c c c c @{}}
\hline
     & \multicolumn{2}{c}{SU Boo} & \multicolumn{2}{c}{RV CVn} \\ 
   Parameter  & Value & Error & Value & Error \\ 
 \hline
  $T_1$ [K]  & \multicolumn{2}{c}{8450 (fixed)} & \multicolumn{2}{c}{6100 (fixed)}\\
  $T_2$ [K]  & 5090  & 60    & 5564 & 75    \\ %
  $i$ [deg]  & 83.18 & 0.14  & 84.75& 0.25  \\ %
  $\Omega_1$ & 5.075 & 0.013 & 3.039& 0.013 \\ %
  $\Omega_2$ & 5.717 & 0.020 & --   & --    \\ %
  $L_1$ [\%] & 94.5  & 0.7   & 50.1 & 0.5   \\ %
  $L_2$ [\%] & 5.5   & 0.2   & 49.9 & 0.5   \\ %
  $r_1/a$    & 0.239 & 0.007 & 0.482& 0.005 \\ %
  $r_2/a$    & 0.184 & 0.006 & 0.481& 0.005 \\ %
 $q=M_2/M_1$ & 0.852 & 0.005 & 0.93 & 0.03  \\ %
 $e$         & \multicolumn{2}{c}{0.00 (fixed)} & \multicolumn{2}{c}{0.00 (fixed)} \\ %
 $F_1=F_2$   & \multicolumn{2}{c}{1.00 (fixed)} & \multicolumn{2}{c}{1.00 (fixed)} \\ %
 $A_1=A_2$   & \multicolumn{2}{c}{1.00 (fixed)} & \multicolumn{2}{c}{0.50 (fixed)} \\ %
 $g_1=g_2$   & \multicolumn{2}{c}{1.00 (fixed)} & \multicolumn{2}{c}{0.32 (fixed)} \\ %
 \hline
\end{tabular}
\end{minipage}
\end{table}

\begin{figure}
  \centering
  \includegraphics[width=0.48\textwidth]{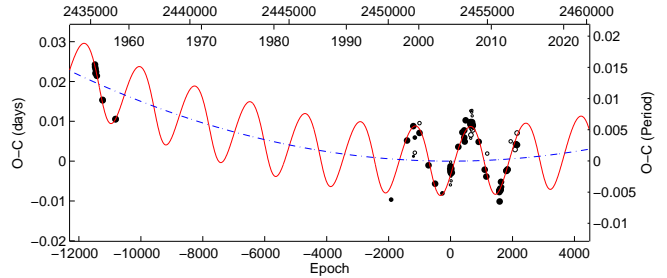}
  \caption{Period analysis of SUBoo. The open circles stand for the secondary, while the filled dots
  for the primary minima. The blue dash-dotted line represents the quadratic term in ephemerides.}
  \label{FigSUBoo_LITE}
\end{figure}

We collected all available times of minima published since its discovery. These are given in Table
\ref{TabMIN}, including our six new measurements. Most of the data points used for the analysis
were derived using the WASP \citep{2006PASP..118.1407P} photometry, the NSVS photometry
\citep{NSVS}, and two also from the CRTS data \citep{2009ApJ...696..870D}. The resulting $O-C$
diagram is plotted in Fig. \ref{FigSUBoo_LITE}, where the periodic variation is clearly visible,
nowadays covering several cycles. We used the same approach as for V418~Aql and the LITE hypothesis
to analyse the period variations. The parameters of the LITE fit are written in Table
\ref{LITEparam}. The period of LITE variation is about only 7.4~years, which makes this system even
more interesting. Moreover, there was detected also a slow steady increase of the period of the
eclipsing pair (see the blue dash-dotted line in Fig. \ref{FigSUBoo_LITE}), probably caused by a
mass transfer between the close components. If we use eq.\ref{masstransfer} we can estimate its
rate to be about 2$\cdot10^{-7}\mathrm{M_\odot}$/yr, which is quite realistic value for a
conservative mass transfer in eclipsing binary.

\begin{figure}
  \centering
  \includegraphics[width=0.48\textwidth]{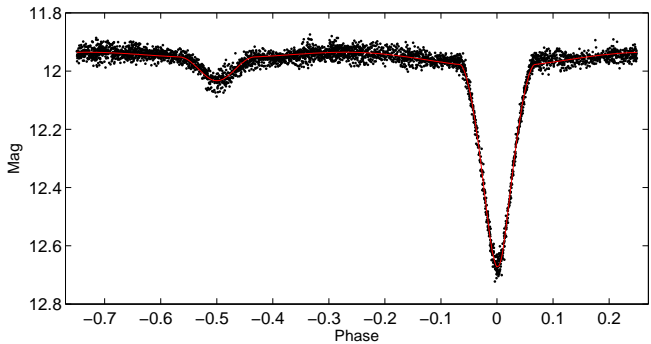}
  \caption{Light curve of SU Boo from the WASP survey and our final fit.}
  \label{FigSUBooLC}
\end{figure}

We used the WASP data for the LC analysis. Despite having no spectroscopy and the radial
velocities, some of the parameters have to be fixed or only estimated. At first, the ephemerides
were fixed according to the period analysis. Secondly, the albedo and gravity darkening values were
kept fixed at their suggested values for stars with radiative envelopes (i.e. $A_i=1$, $g_i=1$,
$i=1,2$). The temperature of the primary component was kept fixed at a value of $T_1 = 8450$~K, in
agreement with its spectral type \citep{1988BAICz..39..329H}. We used the program {\sc PHOEBE},
ver. 0.31a \citep{2005ApJ...628..426P}, which is based on the Wilson-Devinney algorithm
\citep{1971ApJ...166..605W} and its later modifications. For the whole computation process the
eccentricity was fixed at zero. The limb-darkening coefficients were automatically interpolated
from the van Hamme's tables \citep{vanHamme1993}.

During the fitting process, we finally get the answer why there was so large discrepancy between
the two inclination angles published previously. Starting with equal components ($q=1$), the
inclination resulted in $i=81.5^\circ$, while if we fit the mass ratio, it decreases and hence the
inclination increases. We get the smallest possible value of the chi-square when the value of
$q=0.85$ and the inclination about $i=83.2^\circ$. Moreover, the solution presented by
\cite{1960MmSAI..31..107B} is doubtful because of $q > 1$, which is rather improbable. For our
final solution see Fig. \ref{FigSUBooLC} and the parameters of the light curve given in Table
\ref{TabLC}. As one can see, the primary component dominates the system luminosity and also the
mass.

The resulting mass function of the predicted third body (see Table \ref{LITEparam}) yields the
minimal mass of such component of about 1~M$_\odot$ (with the assumption that the orbits are
coplanar and the masses of the eclipsing components are 1.95 and 1.66~$M_\odot$). It is noteworthy
that no additional third light was detected in the LC solution. Assuming a normal main sequence
star, then such a component should contribute about 3\% to the total light, which probably should
be detectable. More precise observations are needed. However, one can also speculate about an
underluminous or even binary nature of the third star. From the estimated luminosities of all
components, the photometric distance to the system was derived to be of about 1.5~kpc.

\subsection{RV CVn}

RV CVn (= AN 4.1921, $V$=14.9~mag) is another seldom investigated eclipsing binary system. It is
the W~UMa-type star, discovered as a variable by \cite{1921AN....214...71L}. Its spectral type was
derived as F8, according to \cite{1927ApJ....65..124S}. In this latter paper,
\citeauthor{1927ApJ....65..124S} stated that the star is of W~UMa type, but no reliable LC solution
was given. Moreover, there was a discussion about its membership to the cluster NGC 5272 (= M 3),
which seems nowadays rather improbable. Another paper by \cite{1981IBVS.1932....1H} also presented
only the light curves, but no LC solution to the data. Since then, many new observations of times
of minima were published, however no reliable LC solution is available till now.

\begin{figure}
  \centering
  \includegraphics[width=0.48\textwidth]{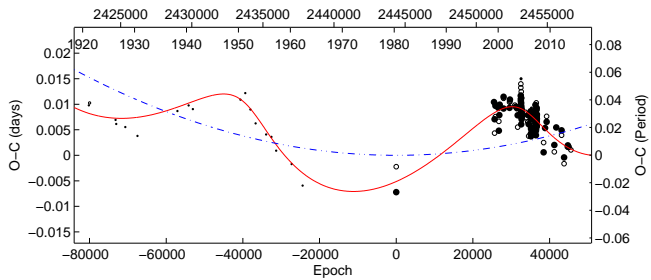}
  \caption{Period analysis of RV CVn.}
  \label{FigRVCVn_LITE}
\end{figure}

\begin{figure}
  \centering
  \includegraphics[width=0.48\textwidth]{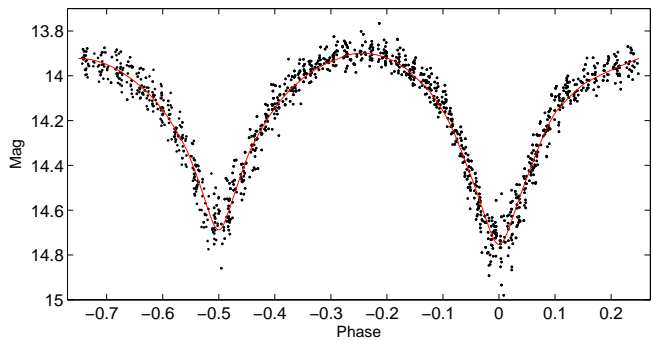}
  \caption{Light curve of RV CVn from the WASP data.}
  \label{FigRVCVn_LC}
\end{figure}

Hence, we collected all the minima observations for a period analysis as well as the data obtained
within the WASP survey project for the LC analysis. For the LC analysis a similar approach as for
SU~Boo was used: the primary temperature was fixed at a value of 6100~K (i.e. spectral type F8),
and the circular orbit. The relevant light curve quantities are given in Table \ref{TabLC}. As one
can see, both the components are similar to each other, the LC fit is plotted in
Fig.\ref{FigRVCVn_LC}. The contact configuration of the system is obvious, as it is usual for these
type of compact W~UMa-type systems.

The set of times of minima is rather huge nowadays: 153 new minima were derived from the WASP
photometry, 6 new from the LINEAR data \citep{2011AJ....142..190S}, while 10 from the CRTS survey.
Eight new observations were obtained by the authors. All of the used minima are given in Appendix
Tables \ref{TabMIN}. The periodic variation is clearly visible, despite rather large scatter of the
older visual or photographic observations (we believe all published data are trustworthy due to
rather deep eclipses, so all of them were used for the analysis). Hence, we followed the same
procedure as for the previous systems and the LITE hypothesis was used. The resulting fit is
plotted in Fig. \ref{FigRVCVn_LITE}, while its parameters are written in Table \ref{LITEparam}.
Besides the 52-yr LITE variation there was also detected a steady period increase. However, it is
rather slow, so the mass transfer between the components is only about
$10^{-7}~\mathrm{M_\odot}$/yr. Such a value is reliable for the contact binary like RV~CVn.

From the resulting mass ratio from the LC analysis ($q=0.93$), and using the assumption of the main
sequence primary component, we can estimate the mass of the secondary. With this mass using the
LITE hypothesis one can also calculate the mass function of the third hidden component, which leads
to the minimal mass (i.e. $i_3 = 90^\circ$) of the third body $M_{3,min} = 0.17~$M$_\odot$.
Therefore, if such body orbits on coplanar orbit with the eclipsing pair, then its light
contribution to the total luminosity of the binary should be negligible. This is also the result of
our LC analysis, where no additional third light was found. Also the interferometric detection is
inapplicable because of its low luminosity.

\subsection{CR Cas}

The system CR Cas (= AN 450.1934, $V$=11.70~mag) was discovered as a variable star by
\cite{1935AN....254..167N}. Later \citep{1939AN....268..165G} was classified as an Algol-type with
the orbital period about 1.42~days, half of the correct value. Its spectral type according to the
SIMBAD database is K8, which is definitely wrong. The precise UBV photometry outside of eclipse
published by \cite{1992AJ....104..801L} was used to derive the unreddened index
$(B-V)_0=-0.27~$mag. Almost the same value of $(B-V)_0$ was derived using the values of Str\"omgren
magnitudes by \cite{1998A.AS..128..139C} (following the method described in
\cite{2001A&A...369.1140H}). This $(B-V)_0$ index corresponds to the spectral type B0.5V-B1V
\citep{1980ARA&A..18..115P}. Later, \cite{1996ApJS..106..133P} gives the type of B. The most
detailed analysis of the star was published by \cite{1998A.AS..128..139C}. They obtained the $uvby$
photometry and the consequent analysis yielded that the components are probably of B0.5V and B1V
spectral types, but located away from the Sun (more than 3.5 kpc, the reddening of the system is
$E(b-y)=0.621$). Moreover, our observations show that the system has total eclipses (lasting about
45 minutes).


\begin{figure}
  \centering
  \includegraphics[width=0.48\textwidth]{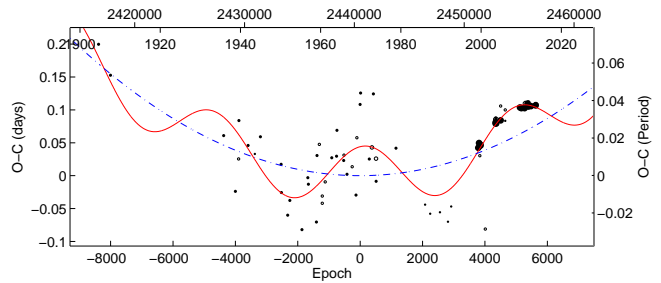}
  \caption{Period analysis of CR Cas.}
  \label{FigCRCas_LITE}
\end{figure}

We collected all available published times of minima for the period analysis. Moreover, 14 new
minima were derived from the NSVS and OMC photometry, and a few other minima were observed by the
authors. All of these data are stored in the Appendix Tables. The same period analysis was used as
in the previous cases, yielding a set of LITE parameters given in Table \ref{LITEparam}. As one can
see from Fig.\ref{FigCRCas_LITE}, the periodic variation is clearly visible nowadays, even despite
rather large scatter of the older visual observations. Moreover, there was also detected rather
rapid period increase (i.e. fitting also the quadratic term in ephemerides).
\cite{1998A.AS..128..139C} speculated about the emission-type secondary component, which probably
should be connected with the rapid mass transfer between the components. From our analysis there
resulted the value of about $3\cdot 10^{-6}~\mathrm{M_\odot}$/yr, which is the largest mass
transfer in our sample. Nevertheless, as noted e.g. by \cite{2001icbs.book.....H}, such a value is
still possible on a  thermal-time scale in binaries. On the other hand, without the detailed
spectroscopic analysis, this is still only a hypothesis.

From the third-body fit, we can also derive the mass function ($f(m_3) = 0.347 \pm
0.015$~~M$_\odot$), from which the minimal mass of the third body resulted in about 6.6~M$_\odot$.
Hence, we can speculate about its detectability in the LC solution performed by
\cite{1998A.AS..128..139C}. The third light fraction resulted in more than 6\% of the total light,
which should be detectable. However, the authors did not test the presence of additional light in
their LC solution. We can also compute the predicted angular separation of the third component
assuming the coplanar orbits and using the photometric distance as derived by
\cite{1998A.AS..128..139C}. This resulted in about $a = 9.4$~mas, which is within the capabilities
of modern stellar interferometers, however its low luminosity makes it probably undetectable with
nowadays facilities.

\subsection{GV Cyg}

The system GV Cyg (= AN 354.1929, $V$=13.2~mag) is the least studied system in our sample. It is
the Algol-type eclipsing binary with the orbital period of about 0.99 day. It was discovered as a
variable by \cite{1930MiSon..17....1H}. The first brief analysis and the LC of the system was
published by \cite{1941KVeBB...6....4A}, which revealed rather deep primary eclipse of about 2
magnitudes, and probably rather shallow secondary. The updated ephemerides were presented by
\cite{1963AJ.....68..257W}, while its spectral type was estimated of about A5 by
\cite{1980AcA....30..501B}. However, no detailed LC and RV analysis exist and the published papers
during the last two decades only contain new times of minima observations.

Hence, we collected all available minima timings, as well as a few our new observations for a
period analysis. Our complete data set consists of more than 70 observations spanning over 80
years. All of the data points are stored in Appendix Tables. As one can see from
Fig.\ref{FigGVCyg1}, the periodic variation is clearly visible especially on the more precise
observations obtained during the last two decades.

The LITE hypothesis applied to the data points led to the parameters presented in Table
\ref{LITEparam}. The period of LITE variation resulted in about 27~years, while the amplitude is
about 11 minutes. Using the very rough parameters of the system as published by
\cite{2004A&A...417..263B}, we can calculate the predicted minimal mass of the third component.
This resulted in about 0.3~M$_\odot$ (hence an M dwarf), which should contribute only negligible
and hardly detectable portion to the total luminosity. Only detailed spectral analysis would detect
such body in the system via spectral disentangling. The quadratic term in ephemerides show some
indication of a slow mass transfer between the eclipsing components, the smallest in our sample,
about only $9\cdot 10^{-9}~\mathrm{M_\odot}$/yr.

\begin{figure}
  \centering
  \includegraphics[width=0.48\textwidth]{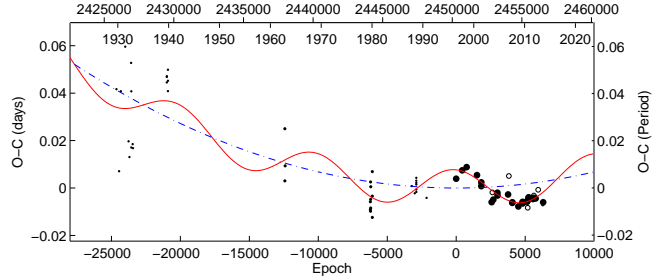}
  \caption{Period analysis of GV Cyg.}
  \label{FigGVCyg1}
\end{figure}


\subsection{V432 Per}

The system V432~Per (= GSC 02856-01647 = TYC 2856-1647-1, $V$=11.2~mag) is probably the most often
studied system in our sample. Its is relatively bright, with short orbital period (about 0.4~day),
has deep minima (about 0.7~mag), and has high declination, which all make it an ideal target for
observers from the northern hemisphere. Its first photoelectric lightcurves in $BV$ filters were
published by \cite{1992IBVS.3797....1A}, later \cite{2002AJ....123..443Y} published the first LC
solution of the system, revealing its asymmetric shape (O'Connell effect) and contact W UMa-type
configuration. More recently, \cite{2008AJ....135.1523L} published their photometric study of the
star, where they presented the spectral types for primary and secondary component as G4 and G8-9.
Moreover, they also found a periodic modulation of minima timings, which led to period about 35~yr,
which could be caused by a hidden M-type component. And finally most recent paper on the star by
\cite{2009MNRAS.400.2085O}, which benefits from a few spectral observations more or less affirms
the results published by \cite{2008AJ....135.1523L}.

\begin{table}
 \begin{minipage}{90mm}
  \caption{Final parameters of the two LITE orbits of V432 Per.}  \label{LITEparamV432Per}
 \centering

  \begin{tabular}{@{}c c@{}}
\hline
 Parameter           &   Value                    \\
 \hline
 $JD_0 $             & 2448601.3757 $\pm$ 0.0020  \\
 $P$ [d]             & 0.38330916 $\pm$ 0.00000013\\
 $p_3$ [day]         & 19125.4 $\pm$ 927.5        \\
 $p_3$ [yr]          & 52.36 $\pm$ 2.54           \\
 $A$  [day]          & 0.0324 $\pm$ 0.0022        \\
 $T_0$               & 2439151.2 $\pm$ 781.3      \\
 $\omega$ [deg]      & 180.9 $\pm$ 14.7           \\
 $e$                 & 0.459 $\pm$ 0.141          \\
 $p_4$ [day]         & 3490.0 $\pm$ 150.9         \\
 $p_4$ [yr]          & 9.55 $\pm$ 0.41            \\
 $A_4$ [day]         & 0.0038 $\pm$ 0.0015        \\
 $T_{0,4}$           & 2451167.5 $\pm$ 182.7      \\
 $\omega_4$ [deg]    & 127.3 $\pm$ 10.4           \\
 $e_4$               & 0.014 $\pm$ 0.008          \\ \hline
 $f(m_3)$ [M$_\odot$]& 0.092 $\pm$ 0.001          \\
 $M_{3,min}$ [M$_\odot$]& 0.81 $\pm$ 0.01         \\
 $f(m_4)$ [M$_\odot$]   & 0.003 $\pm$ 0.001       \\
 $M_{4,min}$ [M$_\odot$]& 0.28 $\pm$ 0.01         \\
 \hline
\end{tabular}
\end{minipage}
\end{table}

\begin{figure}
  \centering
  \includegraphics[width=0.48\textwidth]{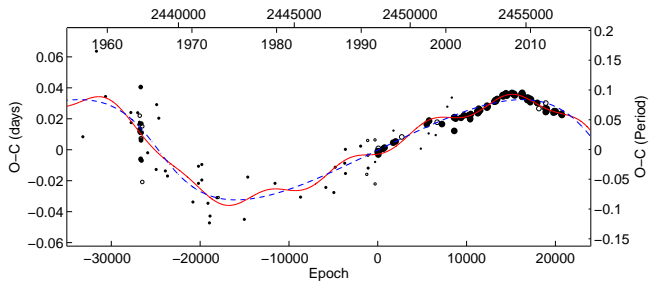}
  \caption{Period analysis of V432 Per. The blue dashed line stands for the third body, while
  the red solid one for the final fit of both LITE terms.}
  \label{FigV432Per._LITE1}
\end{figure}

\begin{figure}
  \centering
  \includegraphics[width=0.48\textwidth]{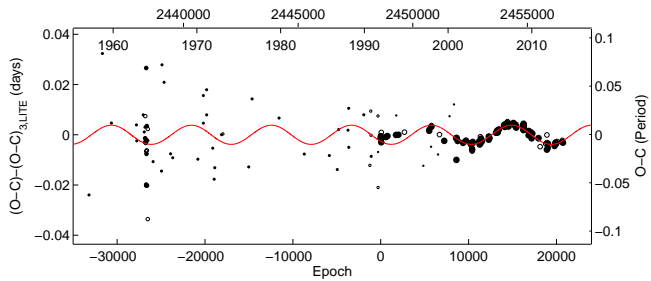}
  \caption{Period analysis of V432 Per, after subtraction of the third-body LITE term.}
  \label{FigV432Per._LITE2}
\end{figure}

Since its discovery, there were published about 200 times of minima. In spite of the fact that the
set of minima is quite large and several studies on period changes were published, we still believe
that its true nature is different than the already published one. The problem with the
interpretation of the $O-C$ diagram is that if we collect all available times of minima and use the
LITE hypothesis as presented in published papers, there still remains an unexplainable variation of
the residuals. Hence, we believe that one has to use two LITE terms for describing the data in
detail.

Therefore, we applied double LITE hypothesis, hence twelve parameters were fitted in total. The
list of all available data found in literature is given in Appendix Tables, while our results are
written in Table \ref{LITEparamV432Per}. The final fit is presented in two Figs.
\ref{FigV432Per._LITE1} and \ref{FigV432Per._LITE2}, where both LITE terms are plotted to the
available data points. Despite rather large scatter of the older visual and photographic minima,
the most recent data obtained during the last decade clearly shows the additional variation
superimposed on the third-body LITE orbit. However, such approach is nothing novel, the double
periodic LITE hypothesis was used for several eclipsing systems, see e.g.
\cite{1996A&AS..120...63B}.

We can only speculate about the nature of these variations. \cite{2008AJ....135.1523L} presented
that the period modulation is most probably caused by the third body orbiting around the EB pair,
and also the third light contribution found in the LC solution originates from this component.
However, in our case the problem is more complicated due to two other components. Their total
luminosity has to be lower than detected $l_3$ in the LC solution presented by
\cite{2008AJ....135.1523L}, however to disentangle their individual contributions is impossible.
Moreover, using the distance to the system as presented in \cite{2008AJ....135.1523L}, we can
compute the predicted angular separation of the third and fourth body in the system for a
prospective interferometric detection. For the third body there resulted the separation about
68~mas, while for the fourth body about 23~mas. The hope of finding these bodies is diminished due
to rather low brightness of the system. Hence, spectroscopic detection via disentangling seems to
be nowadays the best method how to solve this problem.

\subsection{BD +42 2782}

BD +42 2782 (= TYC 3080-1410-1, $V$=9.5~mag) is the brightest star in our sample, and also the
system studied in the most detailed analysis. It was discovered as a variable by \cite{IBVS5600}. A
detailed LC and RV analysis of the star was performed by \cite{2007AJ....133..255L}. They derived
that the system is in contact, its curve is of W~UMa-type, having a cool spot on the primary
component, which is surprisingly underluminous. Its spectrum is probably about F5 and distance was
derived to be of about 124~pc. Moreover, there is also a visual component about
3.5$^{\prime\prime}$ distant, which contributes about 6.5\% to the total light of the system.

\begin{figure}
  \centering
  \includegraphics[width=0.48\textwidth]{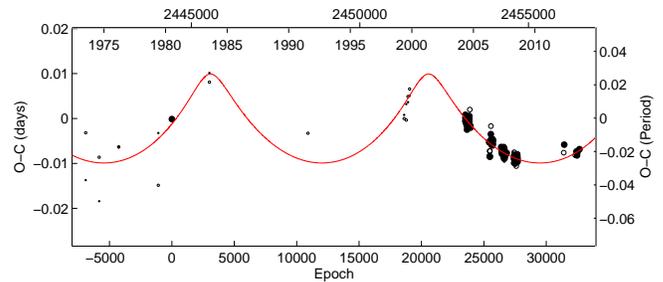}
  \caption{Period analysis of BD +42 2782.}
  \label{FigBD_LITE}
\end{figure}

Despite having a huge set of photometric observations from the WASP survey covering whole LC, we
decided not to perform the LC analysis due to more detailed LC+RV analysis published by
\cite{2007AJ....133..255L} based on precise observations obtained in two filters. However, the WASP
data were used for the minima times derivation and hence the collection of minima (see Table
\ref{TabMIN}) is rather large. Moreover, we derived also new minima from the discovery paper by
\cite{IBVS5600}, from TYCHO \citep{HIP}, from NSVS, and our new recent observations. However, there
arises two major problems, at first the amplitude of photometric variations is about 0.25~mag,
while the precision of individual photometric observations published by \cite{IBVS5600} is
0.01~mag. A similar problem is with the data sampling (it is the sparse photometry, two subsequent
data points are more than 24 minutes separate, which is about 1/20 of the orbital period). All of
these make the scatter of the older observations rather large.

As one can see from Fig. \ref{FigBD_LITE}, the periodic variation is pretty well covered nowadays.
Applying the LITE hypothesis to the data, we get the parameters given in Table \ref{LITEparam}.
There are only two cycles covered with observations, however the variation is evident, mainly
during the last two decades. One can ask how reliable the fit presented in Fig.\ref{FigBD_LITE} is,
but the lack of other observations prevent us to do more for the analysis. The period of LITE was
derived relatively well, but the amplitude should be a bit different because of poor coverage near
both periastron passages.

From the LITE hypothesis there arises that the predicted minimal mass of the third body should be
about 0.5~M$_\odot$, which should be detectable in the LC solution. However, dealing with the
visual component and also this predicted third one, we cannot disentangle the third light into the
contributions from the individual components. Hence we also deal with quadruple system. Using the
distance to the system as derived by \cite{2007AJ....133..255L}, we can also estimate the predicted
angular separation of the third component from the eclipsing pair. This value resulted in about $a
= 78 \pm 10$~mas, which should be easily detectable with nowadays stellar interferometers. However,
there arises a problem with its luminosity. It should be more than 3~mag fainter than the eclipsing
pair (which itself is rather faint for interferometry), so its detection is right on the limit of
nowadays technique.

\section{Conclusion}

We performed the period analysis of times-of-minima for seven rather seldom-investigated eclipsing
systems, where the third component orbiting around the EB pair was suggested as a realistic
hypothesis. For some of these systems the periods are adequately short for the third body to be
discovered via spectroscopic monitoring during several seasons. For some others the third light
contribution to the total light in the LC solution was discussed as a most promising technique of
detection. We also discussed the possibility of interferometric detection of the additional
components, but this was mostly ruled out due to their low luminosities. Moreover, for the
prediction of angular separation of the third components the distance of the systems from the Sun
is needed, which is still the unavailable information for some of the systems.

The most interesting system in our sample is probably CR~Cas, being the most distant and also the
most massive system in our sample. Moreover, besides a proposed third body orbiting around the EB
pair with period about 37~yr, there was derived a rather rapid mass transfer between the eclipsing
components. Hence, the system is maybe in some interesting evolutionary stage. Another noteworthy
system is also SU~Boo, having the shortest period of the third body in our sample about 7.4~yr only
and also the significant mass transfer.

One can also ask whether such periodic variation is presented in a specific kind of EBs, or it is
quite common phenomenon. There arises (e.g. \citealt{2012MNRAS.424.1925C}) that most of the
early-type stars are multiples, hence also the LITE should be detected for many of them. Usually,
there is a problem with an insufficient data set for such an analysis. On the other hand, there are
many EBs observed for decades, where no period variation was detected. Such systems are e.g.
AA~And, AE~Cyg, ER~Vul, and many others. For a catalogue of available $O-C$ diagrams of EBs see
\cite{2006OEJV...23...13P}, or \cite{2001aocd.book.....K}.

The benefit of such period analyses for the stellar multiplicity studies in general is undisputed.
There exist a few hundreds of LITE systems, and their period variation is still being monitored.
Hence, on the longer time scales one can still hope to find some dynamical effects due to the third
bodies. All of our systems are certainly stable (from the ratio of periods), but generally the
orbits of both inner and outer bodies are not stable and are subject of long-term precession, which
can be studied in the future.

 \acknowledgments
An anonymous referee is acknowledged for the useful comments and suggestions that significantly
improved the paper. Based on data from the OMC Archive at LAEFF, pre-processed by ISDC. We do thank
the ASAS, WASP, NSVS, CRTS, and LINEAR teams for making all of the observations easily public
available. We would like to thank Ms.Lenka Kotkov\'a, Mr.Kamil Hornoch, and Mr.Dalibor Han\v{z}l
for obtaining some of the photometric observations. This work was supported by the Czech Science
Foundation grant no. P209/10/0715, by the grant UNCE 12 of the Charles University in Prague, and by
the research programme MSM0021620860 of the Czech Ministry of Education. This research has made use
of the SIMBAD database, operated at CDS, Strasbourg, France, and of NASA's Astrophysics Data System
Bibliographic Services.

\begin{table}
 \centering
 \begin{minipage}{95mm}
\tiny
  \caption{The minima times used for the analysis.}  \label{TabMIN}
\begin{tabular}{@{}l l l l l l @{}}
\hline
 System   & HJD-2400000 &  Error  & Filter & Type  & Reference \\
 \hline
 V418 Aql & 25445.4800  &         &       &  Prim  & \cite{GCVS}               \\ 
 V418 Aql & 25445.5180  &         &       &  Prim  & \cite{KVBB19.33}          \\ 
 V418 Aql & 25492.4200  &         &       &  Prim  & \cite{1966VeSon...7...61G}\\ 
 V418 Aql & 25807.4300  &         &       &  Prim  & \cite{1966VeSon...7...61G}\\ 
 V418 Aql & 25834.4040  &         &       &  Prim  & \cite{KVBB19.33}          \\ 
 V418 Aql & 25854.3800  &         &       &  Prim  & \cite{1966VeSon...7...61G}\\ 
 V418 Aql & 25863.3490  &         &       &  Prim  & \cite{KVBB19.33}          \\
\hline
\end{tabular}
 \begin{list}{}{}
 \item[] Note: W - special filter of the WASP survey
 \end{list}
\end{minipage}
\end{table}

\begin{table}
 \centering
 \begin{minipage}{95mm}
\tiny
  \caption{The minima times used for the analysis.}
  \begin{tabular}{@{}l l l l l l @{}}
\hline
 System   & HJD-2400000 &  Error  & Filter & Type  & Reference \\
 \hline
 SU Boo   & 54159.7399  & 0.00156 &   W   &  Sec   & WASP - This paper         \\ 
 SU Boo   & 54163.6402  & 0.00052 &   W   &  Prim  & WASP - This paper         \\ 
 SU Boo   & 54170.6626  & 0.00095 &   W   &  Prim  & WASP - This paper         \\ 
 SU Boo   & 54185.4972  & 0.0006  &   I   &  Prim  & \cite{2007IBVS.5802....1H}\\ 
 SU Boo   & 54191.7428  & 0.00119 &   W   &  Prim  & WASP - This paper         \\ 
 SU Boo   & 54195.6444  & 0.00160 &   W   &  Sec   & WASP - This paper         \\ 
 SU Boo   & 54202.6711  & 0.00042 &   W   &  Prim  & WASP - This paper         \\
\hline
\end{tabular}
 \begin{list}{}{}
 \item[] Note: W - special filter of the WASP survey, L - special filter of the LINEAR survey
 \end{list}
\end{minipage}
\end{table}

\begin{table}
 \centering
 \begin{minipage}{95mm}
\tiny
  \caption{The minima times used for the analysis.} 
  \begin{tabular}{@{}l l l l l l @{}}
\hline
 System   & HJD-2400000 &  Error  & Filter & Type  & Reference \\
 \hline
 RV CVn   & 53130.61524 & 0.00285 &   W   &  Sec   & WASP - This paper         \\  
 RV CVn   & 53132.49109 & 0.00109 &   W   &  Sec   & WASP - This paper         \\  
 RV CVn   & 53132.62514 & 0.00183 &   W   &  Prim  & WASP - This paper         \\  
 RV CVn   & 53135.46001 & 0.00148 &   W   &  Sec   & WASP - This paper         \\  
 RV CVn   & 53137.47844 & 0.00068 &   W    &   Prim  & WASP - This paper          \\  
 RV CVn   & 53137.61079 & 0.00047 &   W    &   Sec   & WASP - This paper          \\  
 RV CVn   & 53138.41967 & 0.00076 &   W    &   Sec   & WASP - This paper          \\
\hline
\end{tabular}
 \begin{list}{}{}
 \item[] Note: W - special filter of the WASP survey, L - special filter of the LINEAR survey
 \end{list}
\end{minipage}
\end{table}

\begin{table}
 \centering
 \begin{minipage}{95mm}
\tiny
  \caption{The minima times used for the analysis.} 
  \begin{tabular}{@{}l l l l l l @{}}
\hline
 System   & HJD-2400000 &  Error  & Filter & Type  & Reference \\
 \hline
 RV CVn   & 54158.59653 & 0.00059 &   W    &   Prim  & WASP - This paper          \\  
 RV CVn   & 54158.73197 & 0.00089 &   W    &    Sec  & WASP - This paper          \\  
 RV CVn   & 54159.67443 & 0.00066 &   W    &   Prim  & WASP - This paper          \\  
 RV CVn   & 54161.69691 & 0.00169 &   W    &    Sec  & WASP - This paper          \\  
 RV CVn   & 54162.64233 & 0.00076 &   W    &   Prim  & WASP - This paper          \\  
 RV CVn   & 54166.68246 & 0.00163 &   W    &   Prim  & WASP - This paper          \\  
 RV CVn   & 54167.62684 & 0.00077 &   W    &    Sec  & WASP - This paper          \\
\hline
\end{tabular}
 \begin{list}{}{}
 \item[] Note: W - special filter of the WASP survey, L - special filter of the LINEAR survey
 \end{list}
\end{minipage}
\end{table}

\begin{table}
 \centering
 \begin{minipage}{95mm}
\tiny
  \caption{The minima times used for the analysis.} 
  \begin{tabular}{@{}l l l l l l @{}}
\hline
   System    & HJD-2400000 &  Error  & Filter & Type  & Reference \\
 \hline
   RV CVn    & 56175.62512 & 0.00136 &   V    &   Sec  & CRTS  - This paper         \\  
   RV CVn    & 56175.76114 & 0.00029 &   V    &  Prim  & CRTS  - This paper         \\  
   RV CVn    & 56406.37806 & 0.00009 &   R    &   Sec  & This paper                 \\  
   RV CVn    & 56443.71346 & 0.00029 &  BVR   &  Prim  & \cite{2013IBVS.6063....1D} \\  
   RV CVn    & 56497.35703 & 0.00007 &   R    &  Prim  & This paper                 \\  
   RV CVn    & 56657.61426 & 0.00010 &   R    &  Sec   & This paper                 \\  
   CR Cas    & 16711.37    &         &        &  Prim  & \cite{1955PZ.....10..330.} \\
\hline
\end{tabular}
\end{minipage}
\end{table}

\begin{table}
 \centering
 \begin{minipage}{95mm}
\tiny
  \caption{The minima times used for the analysis.} 
  \begin{tabular}{@{}l l l l l l @{}}
\hline
   System    & HJD-2400000 &  Error  & Filter & Type  & Reference \\
 \hline
   GV Cyg    & 27042.438   &         &        &  Prim  & \cite{1941KVeBB...6....4A} \\  
   GV Cyg    & 27061.259   &         &        &  Prim  & \cite{1941KVeBB...6....4A} \\  
   GV Cyg    & 29494.363   &         &        &  Prim  & \cite{1941KVeBB...6....4A} \\  
   GV Cyg    & 29495.356   &         &        &  Prim  & \cite{1941KVeBB...6....4A} \\  
   GV Cyg    & 29496.347   &         &        &  Prim  & \cite{1941KVeBB...6....4A} \\  
   GV Cyg    & 29575.603   &         &        &  Prim  & \cite{1941KVeBB...6....4A} \\  
   GV Cyg    & 29576.589   &         &        &  Prim  & \cite{1941KVeBB...6....4A} \\
\hline
\end{tabular}
 \begin{list}{}{}
 \item[] Note: (*) - http://physics.muni.cz/$\sim$chm/dizina.pdf
 \end{list}
\end{minipage}
\end{table}

\begin{table}
 \centering
 \begin{minipage}{95mm}
\tiny
  \caption{The minima times used for the analysis.} 
  \begin{tabular}{@{}l l l l l l @{}}
\hline
   System    & HJD-2400000 &  Error  & Filter & Type  & Reference \\
 \hline
  V432 Per   & 38440.404   &         &        &  Sec   & \cite{1979VeSon...9..125B} \\  
  V432 Per   & 38670.600   &         &        &  Prim  & \cite{1968AN....290..277H} \\  
  V432 Per   & 39038.566   &         &        &  Prim  & \cite{1979VeSon...9..125B} \\  
  V432 Per   & 39058.540   &         &        &  Prim  & \cite{1979VeSon...9..125B} \\  
  V432 Per   & 39151.292   &         &        &  Prim  & \cite{1979VeSon...9..125B} \\  
  V432 Per   & 39436.440   &         &        &  Prim  & \cite{1968AN....290..277H} \\  
  V432 Per   & 39527.281   &         &        &  Prim  & \cite{1979VeSon...9..125B} \\
\hline
\end{tabular}
\end{minipage}
\end{table}

\begin{table}
 \centering
 \begin{minipage}{95mm}
\tiny
  \caption{The minima times used for the analysis.} 
  \begin{tabular}{@{}l l l l l l @{}}
\hline
   System    & HJD-2400000 &  Error  & Filter & Type  & Reference \\
 \hline
  V432 Per   & 52578.4229  & 0.0002  &   C    &  Sec   & \cite{2003IBVS.5399....1B} \\  
  V432 Per   & 52585.3201  & 0.0003  &  BV    &  Sec   & \cite{2002IBVS.5341....1P} \\  
  V432 Per   & 52588.0021  &         &   V    &  Sec   & \cite{VSB40}               \\  
  V432 Per   & 52588.1943  &         &   V    &  Prim  & \cite{VSB40}               \\  
  V432 Per   & 52617.9042  &         &   V    &  Sec   & \cite{VSB40}               \\  
  V432 Per   & 52618.0953  &         &   V    &  Prim  & \cite{VSB40}               \\  
  V432 Per   & 52897.52960 & 0.0001  &   R    &  Prim  & \cite{2006IBVS.5676....1K} \\
\hline
\end{tabular}
\end{minipage}
\end{table}

\begin{table}
 \centering
 \begin{minipage}{95mm}
\tiny
  \caption{The minima times used for the analysis.} 
  \begin{tabular}{@{}l l l l l l @{}}
\hline
   System    & HJD-2400000 &  Error  & Filter & Type  & Reference \\
 \hline
 BD +42 2782 & 41867.79666 & 0.00287 &   B    &  Prim  & \cite{IBVS5600}            \\  
 BD +42 2782 & 41867.99227 & 0.00625 &   B    &  Sec   & \cite{IBVS5600}            \\  
 BD +42 2782 & 42266.07510 & 0.00217 &   B    &  Prim  & \cite{IBVS5600}            \\  
 BD +42 2782 & 42266.26996 & 0.00272 &   B    &  Sec   & \cite{IBVS5600}            \\  
 BD +42 2782 & 42846.11464 & 0.00676 &   B    &  Prim  & \cite{IBVS5600}            \\  
 BD +42 2782 & 42846.29986 & 0.00577 &   B    &  Sec   & \cite{IBVS5600}            \\  
 BD +42 2782 & 44019.49849 & 0.00308 &   B    &  Prim  & \cite{IBVS5600}            \\
\hline
\end{tabular}
 \begin{list}{}{}
 \item[] Note: W - special filter of the WASP survey
 \end{list}
\end{minipage}
\end{table}

\begin{table}
 \centering
 \begin{minipage}{95mm}
\tiny
  \caption{The minima times used for the analysis.} 
  \begin{tabular}{@{}l l l l l l @{}}
\hline
   System    & HJD-2400000 &  Error  & Filter & Type  & Reference \\
 \hline
 BD +42 2782 & 53205.55329 & 0.00062 &   W    &  Prim  & WASP - This paper          \\  
 BD +42 2782 & 53206.48032 & 0.00348 &   W    &  Sec   & WASP - This paper          \\  
 BD +42 2782 & 53207.40477 & 0.00025 &   W    &  Prim  & WASP - This paper          \\  
 BD +42 2782 & 53208.51424 & 0.00022 &   W    &  Prim  & WASP - This paper          \\  
 BD +42 2782 & 53209.44115 & 0.00069 &   W    &  Sec   & WASP - This paper          \\  
 BD +42 2782 & 53212.40258 & 0.00092 &   W    &  Sec   & WASP - This paper          \\  
 BD +42 2782 & 53217.39913 & 0.00085 &   W    &  Prim  & WASP - This paper          \\
\hline
\end{tabular}
 \begin{list}{}{}
 \item[] Note: W - special filter of the WASP survey
 \end{list}
\end{minipage}
\end{table}

\begin{table}
 \centering
 \begin{minipage}{95mm}
\tiny
  \caption{The minima times used for the analysis.} 
  \begin{tabular}{@{}l l l l l l @{}}
\hline
   System    & HJD-2400000 &  Error  & Filter & Type  & Reference \\
 \hline
 BD +42 2782 & 54254.55660 & 0.00032 &   W    &  Prim  & WASP - This paper          \\  
 BD +42 2782 & 54256.59172 & 0.00009 &   W    &  Sec   & WASP - This paper          \\  
 BD +42 2782 & 54260.47961 & 0.00017 &   W    &  Prim  & WASP - This paper          \\  
 BD +42 2782 & 54261.58945 & 0.00015 &   W    &  Prim  & WASP - This paper          \\  
 BD +42 2782 & 54262.51469 & 0.00025 &   W    &  Sec   & WASP - This paper          \\  
 BD +42 2782 & 54263.44025 & 0.00013 &   W    &  Prim  & WASP - This paper          \\  
 BD +42 2782 & 54263.62409 & 0.00023 &   W    &  Sec   & WASP - This paper          \\
\hline
\end{tabular}
 \begin{list}{}{}
 \item[] Note: W - special filter of the WASP survey
 \end{list}
\end{minipage}
\end{table}

\begin{table}
 \centering
 \begin{minipage}{95mm}
\tiny
  \caption{The minima times used for the analysis.} 
  \begin{tabular}{@{}l l l l l l @{}}
\hline
   System    & HJD-2400000 &  Error  & Filter & Type  & Reference \\
 \hline
 BD +42 2782 & 54626.55765 & 0.00010 &   W    &  Prim  & WASP - This paper          \\  
 BD +42 2782 & 54627.48290 & 0.00013 &   W    &  Sec   & WASP - This paper          \\  
 BD +42 2782 & 54628.40827 & 0.00016 &   W    &  Prim  & WASP - This paper          \\  
 BD +42 2782 & 54628.59306 & 0.00019 &   W    &  Sec   & WASP - This paper          \\  
 BD +42 2782 & 54629.51885 & 0.00023 &   W    &  Prim  & WASP - This paper          \\  
 BD +42 2782 & 54630.44382 & 0.00013 &   W    &  Sec   & WASP - This paper          \\  
 BD +42 2782 & 54630.62916 & 0.00012 &   W    &  Prim  & WASP - This paper          \\
\hline
\end{tabular}
 \begin{list}{}{}
 \item[] Note: W - special filter of the WASP survey
 \end{list}
\end{minipage}
\end{table}


\end{document}